\documentstyle[aps,12pt,epsfig]{revtex}
\draft

\newcommand{\bm}{\bibitem}

\tighten

\begin{document}

\title{Dynamical description of the breakup of one-neutron halo nuclei
$^{11}$Be and $^{19}$C}
\date{\today}
\author{S. Typel$^1$ and R. Shyam$^{1,2}$}
\address { 
$^1$National Superconducting Cyclotron Laboratory, Michigan State University,
East Lansing, Michigan 48824-1321, U.S.A. \\ 
$^2$Saha Institute of Nuclear Physics, Calcutta 700064, India} 
\maketitle

\begin{abstract}
We investigate the breakup of the one-neutron halo nuclei $^{11}$Be and
$^{19}$C within a dynamical model of the continuum excitation of the
projectile. The time evolution of the projectile in coordinate space
is described 
by solving the three-dimensional time dependent Schr\"odinger 
equation, treating the projectile-target
(both Coulomb and nuclear) interaction as a time dependent external
perturbation. The pure Coulomb breakup dominates the relative energy spectra
of the fragments in the peak region, while the nuclear breakup is 
important at higher relative energies. The coherent sum of the two 
contributions provides a good overall description of the experimental
spectra.  Cross sections of the first order
perturbation theory are derived as a limit of our dynamical model.
The dynamical effects are found to be of the order of 10-15$\%$ for the
beam energies in the range of 60 - 80 MeV/nucleon. A comparison of our 
results with those of a post form distorted wave Born approximation 
shows that the magnitudes of the higher order effects are dependent on the
theoretical model. 
\end{abstract}
\pacs{PACS numbers: 24.10.Eq., 25.60.-t, 25.60.Gc, 24.50.+g\\
KEYWORD: Coulomb and nuclear breakup, time dependent Schr\"odinger equation,
finite range DWBA and first order perturbation theory} 

\newpage
\section{INTRODUCTION}

$^{11}$Be and $^{19}$C are examples of one-neutron halo nuclei, where 
the loosely bound valence neutron has a large spatial extension 
with respect to the respective cores (see, e.g., \cite{han87,han95,tan96}
for a recent review). Breakup reactions, in which the valence neutron
is removed from the projectile in its interaction with a target
nucleus, have played a very useful role in probing the structure of such
nuclei \cite{tan88,bla91,fuk91,sac93,nak94,nak99,cha00}. Strongly
forward peaked angular distributions for the neutron \cite{ann90,mar96}
and the narrow widths of the parallel momentum distributions
\cite{ber92,orr95,kel95,baz95,ban95,han96,bau98} of the core
fragments are some of the characteristic features of the breakup reactions
induced by these nuclei, which  provide a clean confirmation of 
their halo structure. In the Serber types of models \cite{han96,ser47},
the breakup cross section is directly related to the momentum space wave
function of the projectile ground state. 

The Coulomb breakup is a significant reaction channel in the scattering of
halo nuclei from a heavy target nucleus (see, e.g., \cite{shy01}). It provides
a convenient way to put constraints on the electric dipole response of
these nuclei \cite{ber88,esb91}. The Coulomb breakup of weakly bound
nuclei can also be used in determining the cross sections of the 
astrophysically interesting radiative capture reactions \cite{bau94}.  

The breakup of the halo nuclei have been investigated 
theoretically by several authors using a number of different approaches
(see, e.g., \cite{cha00} for an extensive list of references). Some of
these models \cite{han96,ann94,bar96} use semiclassical geometrical
concepts (mostly Serber type) to calculate the breakup cross sections.  
A direct breakup model (DBM) (which reduces to the Serber model 
in a particular limit \cite{bau00}) has been formulated within the
framework of the post form distorted wave Born approximation (DWBA)
\cite{bau84,cha00}. However so far only the Coulomb breakup of the
halo nuclei has been investigated within this theory. The 
nuclear breakup, for these cases, has been studied mostly within 
the semiclassical \cite{bon88} and eikonal models \cite{yab92,hen96}.

Breakup reactions of halo nuclei can also be described as the inelastic 
excitation of the projectile from its ground state to the continuum
\cite{ber91,shy99}. In an extensively used theoretical approach,
the corresponding T-matrix is written in terms of the prior form DWBA
\cite{bau84}. For the pure Coulomb breakup, the semiclassical approximation
of this theory is the first order perturbative Alder-Winther theory of
Coulomb excitation \cite{ald75}. It has recently been used to analyze the 
data on the breakup reactions induced by $^{11}$Be \cite{nak94} and $^{19}$C
\cite{nak99}. However, higher order excitation effects 
\cite{esb93,esb95,kid96,mel99,tos00} may be substantial in the breakup
of these nuclei. It has been shown \cite{cha00,ban00} that for the
Coulomb breakup of $^{19}$C, the results of the first order semiclassical
Coulomb excitation theory differ strongly from those of the DBM which
includes the higher order effects.

The general methods in the semiclassical description of the excitation 
process which include higher order effects, are the coupled channel
approaches \cite{can92}, explicit inclusion of successive higher order
terms \cite{typ94}, and the direct numerical integration of the
time dependent Schr\"odinger equation \cite{esb95,kid96,mel99,typ99}.
The last method, in which all the higher order effects in the
relative motion of the breakup fragments are included, provides a
fully dynamical calculation of the projectile excitation caused by
both the Coulomb and the nuclear interactions between the projectile
and the target.

The time dependent Schr\"odinger equation method has been used earlier
\cite{esb95,kid96} to investigate the Coulomb breakup of
$^{11}$Li and $^{11}$Be. However, these calculations have employed a straight
line trajectory for the projectile motion (an approximation valid at higher
beam energies) and have ignored the spins of the particles in the nuclear
models. Furthermore, they are partly perturbative in the sense that 
the first order perturbation theory has been used therein to calculate
the energy distribution of the breakup cross section at larger impact
parameters. This procedure may lead \cite{mel99} to a significantly
larger cross section as compared to that obtained in models where the
partly perturbative approximation is not used.

In this paper, we present calculations for both the Coulomb and the nuclear 
breakup of the one-neutron halo nuclei $^{11}$Be and $^{19}$C within 
the time dependent Schr\"odinger equation method (this will be referred
as the dynamical calculation in the rest of this paper). The spins 
of the particles are included explicitly in the nuclear model.
The Hamiltonian describing the internal
motion of the projectile contains a spin dependent interaction. No 
perturbative approximation has been made in the calculation of the
breakup cross sections. We also calculate first order perturbative
results as a limit of our dynamical model so that a consistent investigation
of the role of the dynamical effect in the breakup cross sections is possible. 
We compare the results of our calculations with those of the 
post form finite range DWBA theory to study the magnitudes of the higher order 
effects calculated in different approaches.

Our manuscript is organized in the following way. In section II, we  
give the details of our formalism. The comparison of our calculations  
with the experimental data and the discussion of   
the results are presented in section III. 
The summary and conclusions of our paper are given in section IV.
 
\section{FORMALISM}

In this section, we describe a general method for solving the time dependent
Schr\"odinger equation, which can be used in problems depending upon a single 
three-dimensional dynamical variable. The potentials involved are assumed
to be local. The time development of the wave function of the relative
motion between the neutron and the core nucleus, $[\Psi({\bf r},t)]$,
is given by the time dependent Schr\"odinger equation
\begin{eqnarray} \label{seq}
 i \hbar \frac{\partial}{\partial t}\Psi({\bf r},t)& =
          & \left[ H_{0}({\bf r})+V({\bf r},t)\right] \Psi({\bf r},t),
\end{eqnarray}
where $H_{0}({\bf r})$ 
is the internal Hamiltonian describing the relative motion
between the valence neutron ($n$) and the core nucleus ($b$). 
It is defined as 
\begin{eqnarray} \label{H0}
H_{0} ({\bf r},t) & = & -\frac{\hbar^{2}}{2\mu_{bn}} \Delta
 + V_{0}({\bf r}),
\end{eqnarray}
where $\mu_{bn}$ is the reduced mass of the $n$ + $b$ system and
${\bf r} = {\bf r}_{n} - {\bf r}_{b}$. 
The potential $V_{0}({\bf r})$ could contain a central and
a spin-orbit term.   In Eq.\  (\ref{seq}),
$V({\bf r},t)$ is the time dependent external field exerted by the 
target on the projectile. In the present application, it is the Coulomb
and nuclear interaction between the target ($A$) and the projectile ($a$).
We write
\begin{eqnarray} \label{Psi}
\Psi({\bf r},t) & = & \sum_c \frac{\psi_c(r,t)}{r}
 {\cal Y}_{J_{c} M_{c}}^{\ell_{c}}({\bf \hat{r}}),
\end{eqnarray}
with
\begin{eqnarray}
  {\cal Y}_{J_{c} M_{c}}^{\ell_{c}}({\bf \hat{r}})
 & = &   \sum_{m_{\ell} m_{s}} 
  (\ell_{c} \:  m_{\ell} \: s \: m_{s}|J_{c} \: M_{c})
  Y_{\ell_c m_{\ell}} ({\bf \hat{r}}) \chi_{s m_{s}},
\end{eqnarray}
where $c$ stands for the individual channels characterized by the quantum 
numbers $J_{c}$, $M_{c}$, and $\ell_{c}$. The orbital angular momentum
($\ell_{c}$) is coupled to the neutron spin ($s=\frac{1}{2}$) 
to give the total angular momentum ($J_{c}$) under the assumption of 
zero spin for the core nucleus.  Substituting Eqs.\ (\ref{H0}) and
(\ref{Psi}) into Eq.\ (\ref{seq}), we can write
a set of coupled equations for the radial wave function 
\begin{eqnarray} \label{rseq}
 i \hbar \frac{\partial}{\partial t}\psi_{c^{\prime}}(r,t) & = &
              \sum_{c} h_{c^{\prime} c} (r,t) \psi_{c}(r,t), 
\end{eqnarray}
where
\begin{eqnarray} \label{hccp}
 h_{c c^{\prime}}(r,t) & = & 
 \left[ -\frac{\hbar^{2}}{2\mu_{bn}}
 \left(\frac{\partial^{2}}{\partial r^{2}} 
 - \frac{\ell_{c}(\ell_{c}+1)}{r^{2}}\right)  
 + V_{0}^{c}(r)
 \right] \delta_{c c^{\prime}} 
 + V^{c c^{\prime}}(r,t) .
\end{eqnarray}
The solution of the radial equation (\ref{rseq}) can be written as
\begin{eqnarray} 
\psi_{c}(r,t) & = & 
  \psi_{c}(r,t_{i}) + \frac{1}{i \hbar}
  \int_{t_{i}}^{t} dt^{\prime} \:
 \sum_{c^{\prime}} h_{c c^{\prime}}(r,t^{\prime})
                  \psi_{c^{\prime}}(r,t^{\prime}),
\end{eqnarray}
with the starting time $t_{i}$. Alternatively, the wave function
at time $t$ can be calculated from the initial wave function
by applying the unitary time evolution operator (in the 
obvious matrix notation)
\begin{eqnarray} \label{upsi}
\psi(t) & = & {\bf U}(t_{i}, t) \psi(t_{i}).
\end{eqnarray}
For a small time step $\Delta t$,
the time evolution operator can be approximated as  
\begin{eqnarray} \label{teo}
 {\bf U}(t,t+\Delta t) & \approx & 
 \frac{{\bf 1}+ \frac{\Delta t}{2i\hbar} {\bf h}(t)}{{\bf 1}
  -\frac{\Delta t}{2i\hbar} {\bf h}(t)},
\end{eqnarray}
where the elements of the matrix ${\bf h}$ are given by $h_{c c^{\prime}}$.
Eqs.\ (\ref{upsi}) and (\ref{teo}) 
are the starting points for the numerical solution of the problem,
which has been done by following the method described in Ref.\
\cite{typ99}. This approach has the virtue that a first order calculation
can also be performed within the same program by 
keeping the coupling only between the initial channel ($c^{\prime}=c_{in}$)
and all the possible final channels ($c$) in the
perturbation potential, i.e., by replacing $V^{cc^{\prime}}$ by
$V^{cc^{\prime}}\delta_{c^\prime c_{in}}$ 
in Eq.\ (\ref{hccp}). Then the matrix  ${\bf h}$ is no longer hermitian and 
the time evolution operator ${\bf U}$ becomes non-unitary.

Unlike the previous calculations \cite{esb95,kid96,mel99}, we do not assume
the projectile to move on a straight line trajectory. In our case,
the c.m.\ of the projectile is supposed to follow a hyperbolic
trajectory with respect to the target during the scattering process.
For application to the breakup of the one-neutron halo nuclei,
the Coulomb part of the external perturbation $[V_{C}({\bf r},t)]$
is given by
\begin{eqnarray} \label{vcoul}
 V_{C}({\bf r},t) & = & 
  \frac{Z_{A} Z_{b} e^{2}}{|{\bf r}_{b} - {\bf R}_{A}(t)|} 
  -  \frac{Z_{A} Z_{b} e^{2}}{|{\bf R}_{A}(t)|},
\end{eqnarray} 
where $Z_{A}$ and $Z_{b}$ are the charge numbers of the target and
the core nuclei, respectively. ${\bf R}_A(t)$ is the coordinate of
the target in the projectile centre of mass (c.m.) frame, and
${\bf r}_{b} = - \frac{m_{n}}{m_{n}+m_{b}}{\bf r}$ and 
${\bf r}_{n} = \frac{m_{b}}{m_{n}+m_{b}} {\bf r}$, are 
the position vectors of the core nucleus and the neutron, respectively.
The nuclear part of the external perturbation $[V_{N}({\bf r},t)]$
is the sum of the neutron-target and core-target 
optical model potentials, respectively. It is given by 
\begin{eqnarray}
 V_{N}({\bf r},t) & = & 
 V_{n} f(|{\bf r}_{n}- {\bf R}_{A}(t)|,R_{Vn}, a_{Vn})
 + i W_{n} f(|{\bf r}_{n}- {\bf R}_{A}(t)|,R_{Wn}, a_{Wn})
 \nonumber \\  
 & + & V_{b} f(|{\bf r}_{b}- {\bf R}_{A}(t)|,R_{Vb}, a_{Vb})
 + i W_{b} f(|{\bf r}_{b}- {\bf R}_{A}(t)|,R_{Wb}, a_{Wb}),
\end{eqnarray}
where
\begin{eqnarray}
 f(x,R,a) & = & 
 \left[ 1 + \exp\left( \frac{x-R}{a}\right) \right]^{-1} 
\end{eqnarray}
with the diffuseness parameter $a$ and radius $R$. The depths of the real
and the imaginary parts of the neutron-target and the core-target optical
potentials are denoted by $V_{n}$, $W_{n}$, and $V_{b}$, $W_{b}$, respectively.
The corresponding radii and the diffuseness parameters are given by
$R_{Vn}$, $R_{Wn}$, and $a_{Vn}$, $a_{Wn}$, and $R_{Vb}$, $R_{Wb}$ 
and $a_{Vb}$, $a_{Wb}$, respectively.

Using a multipole expansion of the perturbation potentials,
we can write 
\begin{eqnarray} \label{permul}
 V^{c c^{\prime}}(r,t) & = & \sum_{\lambda \mu} 
 C_{c c^{\prime}}^{\lambda \mu}
 \left[ v_{C}^{\lambda \mu}(r,t) + v_{N}^{\lambda \mu}(r,t)
  \right], 
\end{eqnarray}
with the coefficients
\begin{eqnarray}
 C_{c c^{\prime}}^{\lambda \mu} & = & 
 \int d\Omega \:
 \left[ {\cal Y}_{J_{c} M_{c}}^{\ell_{c}} \right]^{\dagger}
 Y_{\lambda \mu} 
 {\cal Y}_{J_{c^{\prime}} M_{c^{\prime}}}^{\ell_{c^{\prime}}} .
\end{eqnarray}
The time dependent radial potentials are given by
\begin{eqnarray}
 v_{C}^{\lambda \mu}(r,t) & = & 
 \frac{4 \pi Z_{A} Z_{b} e^{2}}{2\lambda +1} 
 \left( \frac{-m_{n}}{m_{n}+m_{b}} \right)^{\lambda}
 \frac{r^{\lambda}}{R_{A}^{\lambda+1}(t)} 
 Y_{\lambda \mu}^{\ast} ({\bf \hat{R}}_{A}(t))
\end{eqnarray}
and 
\begin{eqnarray}
 v_{N}^{\lambda \mu}(r,t) & = & 
 V_{n} f^{\lambda \mu}(r_{n}, R_{Vn}, a_{Vn}, t)
 + i W_{n} f^{\lambda \mu}(r_{n}, R_{Wn}, a_{Wn}, t)
 \nonumber \\  
 & + & V_{b} f^{\lambda \mu}(r_{b}, R_{Vb}, a_{Vb}, t)
 + i W_{b} f^{\lambda \mu}(r_{b}, R_{Wb}, a_{Wb}, t),
\end{eqnarray}
where
\begin{eqnarray}
 f^{\lambda \mu}(r,R,a,t) & = & \int d\Omega  \:
 f(|{\bf r}- {\bf R}_{A}(t)|,R,a) \:
 Y_{\lambda \mu}^{\ast}({\bf \hat{r}}).
\end{eqnarray}

The triple differential cross section for the breakup of the
projectile $a$ into $b$ and $n$ is given by
\begin{eqnarray} \label{d3s}
\frac{d^{3} \sigma}{dE_{bn}d\Omega_{bn}d\Omega_{bn-A}} & = &
     \frac{d \sigma_{R}}{d\Omega_{bn-A}} \: 
  P_{if}({\bf k_{bn}}) \: \rho_{f}(k_{bn}),
\end{eqnarray}
where $\Omega_{bn-A}$ are the angles associated with the relative motion
of the c.m.\ of the $b+n$ system with respect to the target 
nucleus $A$. The first term on the right hand side of Eq.\ (\ref{d3s}) 
is the Rutherford cross section for the projectile-target scattering.
The last term,
\begin{equation}
 \rho_{f}= \frac{\mu_{bn}k_{bn}}{(2\pi)^{3}\hbar^{2}},
\end{equation}
is the density of final states. The excitation probability, $P_{if}$
(which is a function the relative momentum 
$\hbar {\bf k}_{bn}$ between the two fragments), is given by
\begin{eqnarray} \label{Pfi}
 P_{if}({\bf k}_{bn}) & = & \lim_{t_{f} \to \infty}
 \frac{1}{2J_{i} + 1} \sum_{M_{i} m_{s}}
 \left|\langle \Phi_{m_{s}}^{(-)}({\bf k}_{bn},{\bf r})|
 \Psi_{M_{i}}({\bf r},t_{f}) \rangle \right|^2,
\end{eqnarray}
where $\Phi_{m_{s}}^{(-)}({\bf k}_{bn},{\bf r})$ is a complete scattering
solution for the relative motion of the two fragments with the
in-going wave boundary condition.  It satisfies the Schr\"{o}dinger
equation (\ref{seq}) for a vanishing perturbation potential $V({\bf r},t)$.
The wave function $\Psi_{M_{i}}({\bf r},t_{f})$ 
is the solution of the time-dependent
Schr\"odinger equation with the boundary condition
that as $t_{i} \to  -\infty$, it goes to the unperturbed wave
function $\Phi_{M_{i}}$ of the ground state of the projectile.
Since we are interested only in the breakup contributions,
it is advantageous for 
numerical reasons to replace the full wave function 
$\Psi_{M_{i}}({\bf r},t)$, in Eq.\ (\ref{Pfi}),
by the corresponding continuum wave function
\begin{eqnarray}
\Psi_{M_{i}}^{\rm cont}({\bf r},t) 
 & = & \Psi_{M_{i}}({\bf r},t) 
 - \sum_{b^\prime} \langle \Phi_{b^\prime}({\bf r},t)|
            \Psi_{M_{i}}({\bf r},t)\rangle \Phi_{b^\prime}({\bf r},t),
\end{eqnarray}
where the sum runs over all bound states of the system.

In the actual numerical realization, the radial wave functions
$\psi_{c}(r,t)$ are discretized on a mesh with points
$x_{n}= n \Delta x$ where $\Delta x = 0.0025$ and $n=0,\dots,400$.
They are related to the points in the radial coordinate $r$ by
the mapping, $r_{n} = R_{max} [\exp(ax_{n}) - 1]/[\exp(a)-1]$,
with $R_{max} = 900$~fm. The parameter $a$ is chosen so that
$r_{1} = 0.3$~fm. The second order derivative, 
$\frac{\partial^{2}}{\partial r^{2}}$, in Eq.\ (\ref{hccp}) is 
represented by a finite difference approximation. In the first step,
the ground state radial wave function is calculated assuming
a Woods-Saxon potential, $V_{0}^{c}(r) = 
V_{0}f(r,R_{0},a_{0})\delta_{cc^{\prime}}$, for the
$n - b$ interaction in the initial channel with a given
set of quantum numbers, $J_{c}$, $M_{c}$, and $\ell_{c}$. 
The time evolution of this initial wave function is calculated by
the repeated application of the time evolution operator 
(\ref{teo}), using time steps
$\Delta t = 1$~fm/c, and taking into account partial waves with
$\ell = 0,1,2,3$ for the $n - b$ relative motion. 
This leads to a set of coupled linear equations for the radial
wave functions which are 
solved with the technique as described in Ref.\ \cite{typ99}.
We use a time interval for the evolution which is symmetric to the
time of closest approach between the target and the projectile.
Its limits are determined by the condition that the perturbation 
potential is at least 200 times smaller than its maximum value.
In order to avoid spurious excitations, the time dependent potential
was switched on adiabatically. Furthermore, care has been taken
to ensure that unphysical bound states are not populated
during the time evolution.
\begin{table}[here]
\caption{Optical potential parameters for the neutron-target and
core-target interaction.}
\begin{tabular}{lddddddl}
 projectile  & $V_{n}$ [fm] & $R_{Vn}$ [fm] & $a_{Vn}$ [fm] 
             & $W_{n}$ [fm] & $R_{Wn}$ [fm] & $a_{Wn}$ [fm] & Ref. \\
 \hline
 ${}^{11}$B & $-$27.88  & 6.93 & 0.75 & $-$14.28 & 7.47 & 0.58 & 
 \protect\cite{per76} \\
 ${}^{19}$C  & $-$29.48  & 6.93 & 0.75 & $-$13.18 & 7.47 & 0.58 & 
 \protect\cite{per76} \\
 \hline \hline
 projectile  & $V_{b}$ [fm] & $R_{Vb}$ [fm] & $a_{Vb}$ [fm] 
             & $W_{b}$ [fm] & $R_{Wb}$ [fm] & $a_{Wb}$ [fm] & Ref. \\
 \hline
 ${}^{11}$Be & $-$70.0   & 5.45 & 1.04 & $-$58.9  & 5.27 & 0.887& 
 \protect\cite{bon85} \\
 ${}^{19}$C  & $-$200.0  & 5.39 & 0.90 & $-$76.2  & 6.58 & 0.38 & 
 \protect\cite{buexx}
\end{tabular}
\end{table}
\noindent
The final wave function is projected
onto scattering states and the excitation probability is calculated
according to Eq.\ (\ref{Pfi}), after removing the bound state contributions.
For each $M_{c}$ substate of the initial state, an independent
calculation has been performed.
\section{RESULTS AND DISCUSSIONS}

The parameters of the nuclear optical potentials
for the $n$-target and core-target interaction
[see, Eq.\ (16)] used in our calculations, are given in Table I. 
We have adopted a single particle potential model
to calculate the ground state wave function of the projectile. The  
ground state of $^{11}$Be was assumed to have a 2$s_{1/2}$ valence
neutron coupled to the 0${}^{+}$ $^{10}$Be core 
with a binding energy of 504 keV. 
The corresponding single particle wave function was constructed by
assuming the neutron - $^{10}$Be interaction of the Woods - Saxon type
having a central and a spin-orbit part. The radius and the diffuseness
parameters in both the terms were taken to be 2.478~fm and 0.5~fm,
respectively. The depth of the central term was searched so as to 
reproduce the ground state binding energy. 
This 2$s_{1/2}$ wave function has an additional 
node as compared to a simpler zero-range wave function.
The strength of the spin-orbit term
was adjusted by requiring that the first excited ($\frac{1}{2}^{-}$) 
state in ${}^{11}$Be is at the experimental excitation energy of
320~keV.  In the dynamical calculation, we assume a
vanishing $n$-core potential for
the higher partial waves so that unphysical resonances can be
avoided.  For $^{19}$C, the ground state wave function was obtained
with the similar procedure by assuming a configuration in which a
2$s_{1/2}$ neutron is coupled to the 0$^+$ $^{18}$C core. The radius
and the diffuseness parameters associated with the Woods-Saxon 
interaction were taken to be 3.30~fm and 0.65~fm, respectively.
In the dynamical calculations, we took into account only the $n$-core
interaction in the s-wave. The spectroscopic factors for the ground 
\begin{figure}[here]
\begin{center}
\mbox{\epsfig{file=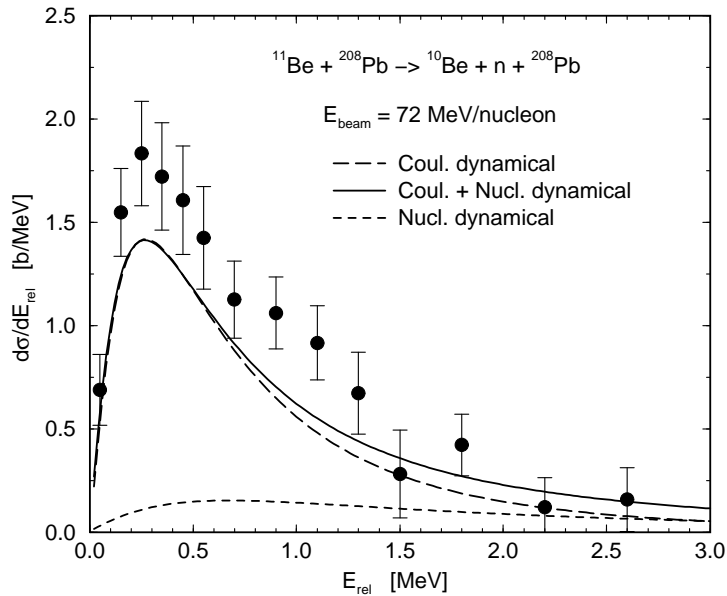,height=8.0cm}}
\end{center}
\caption {
The differential cross section as a function of the relative energy of the
fragments (neutron and $^{10}$Be) emitted in the $^{11}$Be induced
breakup reaction on a $^{208}$Pb target at the beam energy of
72 MeV/nucleon. The long-dashed
and short-dashed lines represent the pure Coulomb and pure nuclear breakup
contributions, respectively. Their coherent sum is depicted by the 
solid line. The experimental data are taken from ~\protect\cite{nak94}. 
}
\label{fig:figa}
\end{figure}
\noindent
state were taken to be 1 in all the cases. In the multipole expansion
of the perturbation potential [Eq. (\ref{permul})], we included contributions
up to $\lambda=2$.

In Fig.\ 1, we compare the results of our calculations with the experimental
data (taken
from \cite{nak94}) for the relative energy spectrum of the fragments
emitted in the breakup reaction of $^{11}$Be on a $^{208}$Pb target at
the beam energy of 72 MeV/nucleon. In these calculations, the integration
over the theta ($\theta_{bn-A}$) angles of the projectile c.m. was done
in the range of $0^{\circ} - 3^{\circ}$, which corresponds to a minimum
impact parameter of about 12~fm. The long-dashed and short-dashed lines
represent the results of the dynamical calculations for the pure Coulomb
and the pure nuclear breakup, respectively, while their coherent sum
is represented by the solid line. The Coulomb cross sections consist mostly of
the dipole term as the contributions of the quadrupole mode are negligible.

We note that the pure
Coulomb contributions dominate the cross sections around the peak value
while the nuclear breakup is important at the larger relative energies. 
This can be understood from the energy dependence of the two contributions.
The nuclear breakup occurs when the projectile and the target nuclei are 
close to each other. Its magnitude which is determined mostly by the
geometrical conditions, has a weak dependence on the relative energy of the
outgoing fragments beyond a certain minimum value. Contrary to this,
the Coulomb breakup contribution has a long range and it shows a strong energy
dependence. The number of virtual photons increases for small excitation
energies. At the same time, a much larger 
range in the impact parameter implies a substantial breakup probability. 
The results shown  in this figure are  
in agreement with those of Ref.\ \cite{das99}. The domination
of the nuclear breakup may explain the failure of the pure Coulomb
finite range DWBA calculations \cite{cha00} in explaining the data at larger 
relative energies.

The coherent sum of the Coulomb and the nuclear contributions provides a
good overall description of the experimental data. Although, the pure
nuclear breakup contributions below 0.5~MeV are substantial, yet their 
interference with the Coulomb part does not change appreciably 
the shape of the total cross section in the peak region. It may be noted  
that the absolute magnitude of the cross section near the maximum
is somewhat underestimated by our calculations, but 
the position of the peak is well reproduced. This is in agreement
with the results of the dynamical calculations reported in \cite{mel99}.
However, in the semiclassical coupled-channel calculations \cite{das99}
for this reaction, the peak position of the calculated
cross sections are shifted towards larger energies as compared to
that of the data. 

It may be remarked that use of a some what smaller minimum impact
parameter in the angular integration over the $^{11}$Be c.m. would increase
the cross sections which may reduce the difference between the
experiment and the theory. However, this would mainly affect the nuclear
contribution since it is more sensitive to the smaller projectile-target
distances. The Coulomb contributions arise mostly from the much larger
impact parameters as long as the excitation energy is not too large.
At very small impact parameters the absorption due to the imaginary
parts of the optical potentials will result in an effective cutoff.
Since the optical potentials are not sufficiently well known,
the choice of the maximum scattering angle introduces an uncertainty
but our choice for this is a reasonable one.

In Fig.\ 2, we show the results of our dynamical calculations for the
relative energy
spectrum of the fragments (neutron and $^{18}$C)  
emitted in the breakup of $^{19}$C on a $^{208}$Pb target at the 
beam energy of 67 MeV/nucleon. The integrations over $\theta_{bn-A}$ 
were done in the range of $0^{\circ} - 3^{\circ}$. Since the
binding energy (B$_{n-core}$)
of the valence neutron - core system in the ground state of $^{19}$C is
still an unsettled issue \cite{shy01,mad01,wou88,orr91,wap93}, we
present in this figure the results of calculations performed with two
values [530 keV (part a) and 650 keV (part b)] of B$_{n-core}$.
This is for the first time that the dynamical calculations
(including the nuclear breakup) have been performed for this case. 
The experimental data are taken from \cite{nak99}. We see that the 
Coulomb breakup dominates the cross sections in the peak region while
the nuclear breakup is important at the larger relative energies in this
case too.
\begin{figure}[here]
\begin{center}
\mbox{\epsfig{file=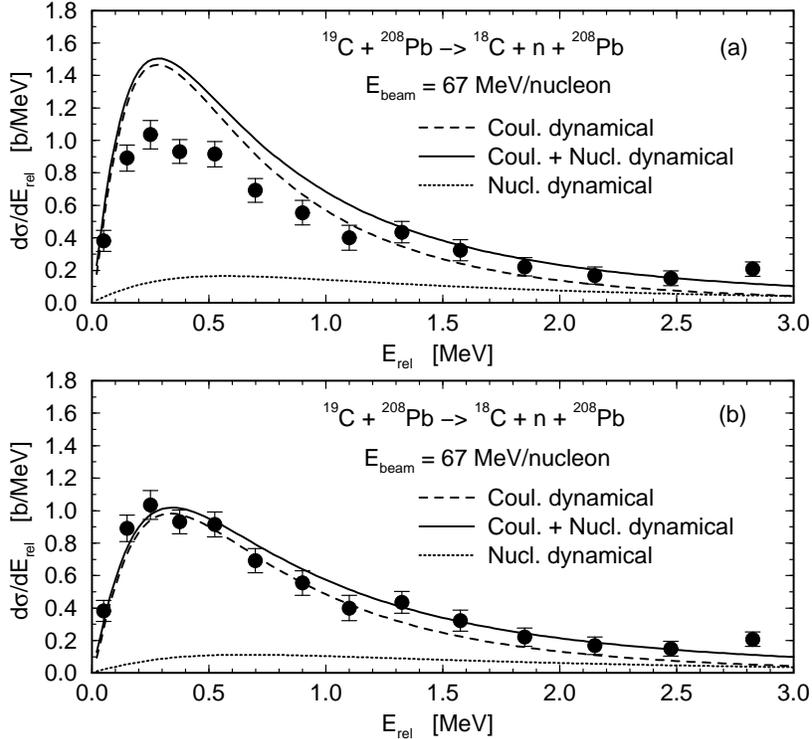,height=10.0cm}}
\end{center}
\caption {
The relative energy spectrum of the fragments (neutron and $^{18}$C) in the
breakup of $^{19}$C on a $^{208}$Pb target at the beam energy of 
67 MeV/nucleon. Part (a) shows the results obtained with a binding
energy of 530 keV for the neutron-core configuration in the ground state
of $^{19}$C while part (b) is the result with a value of 650 keV for 
the same. The dashed and dotted lines represent the pure Coulomb and pure
nuclear breakup contributions, respectively, while their coherent sum is
shown by the solid line. The experimental data are taken from
~\protect\cite{nak99}.
}
\label{fig:figb}
\end{figure}
The dynamical calculations carried out with the binding
energy of 530 keV [part (a)] overestimate the cross sections in the
peak region by about 35 - 40$\%$, while those done with 650 keV 
[part (b)] are in good agreement with the data. Our calculations,
therefore, seem to support the latter value for B$_{n-core}$ which is
within the error range of the value $[(530\pm 130)$~keV] 
reported in \cite{nak99}.  The expected shift in
the peak position in part (b) is within statistical
error of the data. We would like to recall, however, that we have used
a spectroscopic factor of 1 for the neutron-core configuration for the
ground state of $^{19}$C. 

An alternative scenario has been presented in \cite{nak99}, where
the same data have been analyzed within the first order semiclassical
perturbation theory of the Coulomb excitation.  With a binding energy of
530 keV and the same neutron-core configuration for the ground state of
$^{19}$C, these calculations overestimate the data in the peak
region by about 40-50$\%$ (see also \cite{ban00}). Instead of using a
larger value of B$_{n-core}$ in order to fit the data, these authors
try to extract a spectroscopic factor (SF) for this configuration
by comparing their pure Coulomb dissociation calculations with the
data (which are corrected for the nuclear breakup effects in an
approximate way). The value of SF (0.67) obtained in this way
is close to the shell model results of Ref.\ \cite{war92}.

However,
a note of caution must be added about the significance to this result.
The spectroscopic factors reported in \cite{war92} correspond to various
specific states of the $^{18}$C core. The data of Ref. \cite{nak99} are,
however, inclusive in the sense that the specific core states of $^{18}$C
are not observed there. Therefore, measurements of the relative energy 
spectra in experiments of the type reported in \cite{mad01}, where
specific core states of $^{18}$C are identified by tagging to the
decay photons, are required for a meaningful comparison of the
SF extracted from such studies with those of the shell model.  
Moreover, the procedure adopted in \cite{nak99} to correct the data for the
nuclear breakup effects may not be valid due to the long range 
of the nuclear interaction in case of the halo nuclei \cite{das99}.
Obviously, the value of the SF extracted from the breakup studies 
depends critically on the theory used to calculate the corresponding 
cross sections. Therefore, more experimental data and their theoretical
analysis within different models of the breakup reactions are required
for arriving at a more definite conclusion in this regard.
\begin{figure}[h]
\begin{center}
\mbox{\epsfig{file=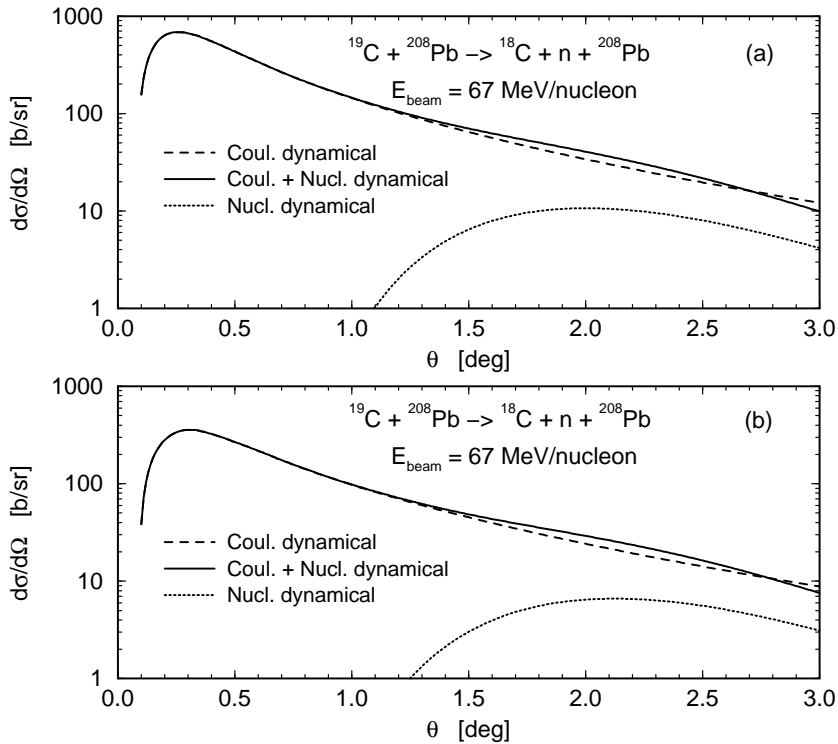,height=10.0cm}}
\end{center}
\caption{
Dynamical model results for the angular distribution of the center
of mass of $n$ + $^{18}$C system in the breakup of $^{19}$C on a Pb
target at the beam energy of 67 MeV/nucleon for two values [530 keV (part a)
and 650 keV (part b)] of the $^{19}$C ground state. The dashed and dotted
lines represent the pure Coulomb and pure nuclear contributions. 
Their coherent sum is represented by the solid line. 
}
\label{fig:figc}
\end{figure}
Another argument put forward in \cite{nak99} in favor of their procedure
is that with the values 0.530 MeV and 0.67 for B$_{n-core}$ and SF,
respectively, the angular distribution of the
$n$ + $^{18}$C c.m.\ measured in the same reaction is well reproduced.
However, this conclusion assumes that the shape of this
angular distribution is not affected by the nuclear breakup effects below 
the grazing angle ($\sim$ 2.7$^\circ$). To study the role of the
nuclear breakup  for this data, we show in Fig.\ 3 the angular
distribution of the 
c.m.\ of the $n$ + $^{18}$C system for the values of B$_{n-core}$ of 
530 keV (upper part) and 650 keV (lower part). The integrations over the
relative energy is performed in the range of 0.0--0.5 MeV (the 
same as done in Ref. \cite{nak99}). 
It is clear from this figure that the nuclear breakup effects start
becoming important already from c.m.\ angles of 1.5$^\circ$,
and the total cross section (coherent sum of 
the Coulomb and nuclear contributions) differs from the pure Coulomb one
even below the grazing angle. Beyond this angle,
the absorptive part of the optical potentials reduces the cross section.
In this figure we have not shown the comparison of our calculations with
the experimental data as it would require folding the calculations with
the experimental angular resolution \cite{nak99}. However, the  
comparison of the unfolded and folded results as shown in \cite{ban00}
suggests that the quality of agreement between calculations done with both the 
options and the data would be similar.

The magnitude of higher order effects in the Coulomb breakup
reactions of $^{11}$Be and $^{19}$C is a subject of current interest.
In an earlier calculation \cite{esb95}, higher order effects were
found to be rather small for the reaction studied in Fig.\ 1. However,
comparing the result of the adiabatic model of Coulomb breakup
reactions with that of the first order semiclassical
perturbation theory of the Coulomb excitation, it has been concluded
in \cite{tos00} that the higher order effects are substantial for the
reaction investigated in Fig.\ 2, which
\begin{figure}[here]
\begin{center}
\mbox{\epsfig{file=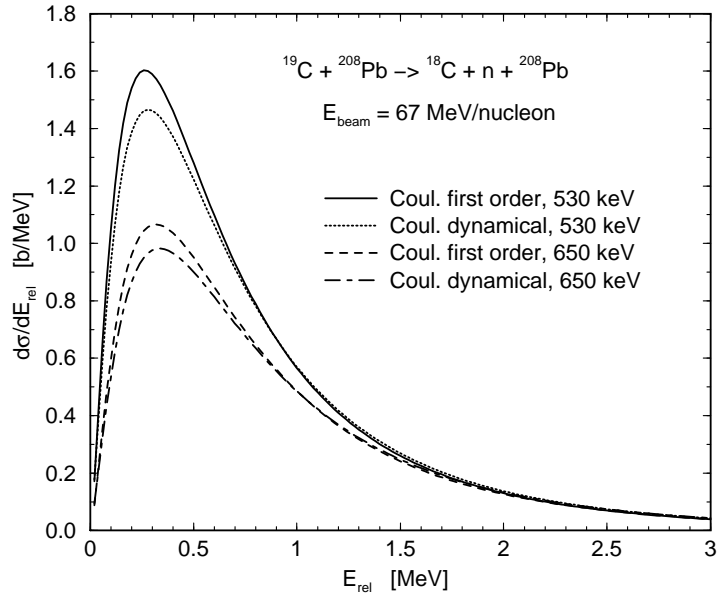,height=8.0cm}}
\end{center}
\caption{
Comparison of the dynamical model and the first order perturbation theory 
results for the pure Coulomb breakup contribution to the relative energy
spectrum of the fragments (neutron and $^{18}$C) emitted in the breakup of
$^{19}$C on a $^{208}$Pb target at the beam energy of 67 MeV/nucleon. 
The solid and dotted lines represent the results of the first order and
the dynamical model calculations, respectively, obtained
with the value of B$_{n-core}$ equal to 530 keV while the dashed
and dashed-dotted lines represent the same for the B$_{n-core}$ value of
650 keV.  The first order perturbation theory results have been obtained  
from the dynamical model in a particular limit as discussed in section II. 
}
\label{fig:figd}
\end{figure}
\noindent
would have a considerable
influence on the extracted spectroscopic factor. In contrast, these
effects were found to be rather small in \cite{typ01}
where the first order and the higher order terms were calculated within
the same model.  As discussed in section II, the first order perturbation 
theory results can be obtained from our dynamical model in a
particular limit.  In Fig.\ 4, the first order results obtained in this way
are compared with those of the full dynamical model for the pure
Coulomb breakup contribution to the same reaction as in Fig. 2. 
It can be noted that the higher order effects are rather small. They reduce
the peak cross sections of the first order theory by about 10$\%$ for 
both values of B$_{n-core}$ but leave the shapes of the spectra largely
unaffected.

In Fig.\ 5, we compare the  
pure Coulomb breakup contributions (calculated within the 
dynamical model and the finite range  DWBA \cite{cha00}) to 
the reactions studied in Figs.\ 1 and 2.  
In the finite range DWBA theory the higher order effects in
the target-fragment interaction are automatically incorporated. The
interesting aspect of this comparison is that while for the $^{11}$Be case
the finite range DWBA results are slightly larger than those of the dynamical 
semiclassical calculation, the former is smaller than the latter for
$^{19}$C induced reaction for both the values of the binding energy. This
clearly shows that higher order calculations for the Coulomb
breakup performed within different theories could be different from
each other. The reason for this difference is the fact that fully
quantal higher order approaches
\begin{figure}[here]
\begin{center}
\mbox{\epsfig{file=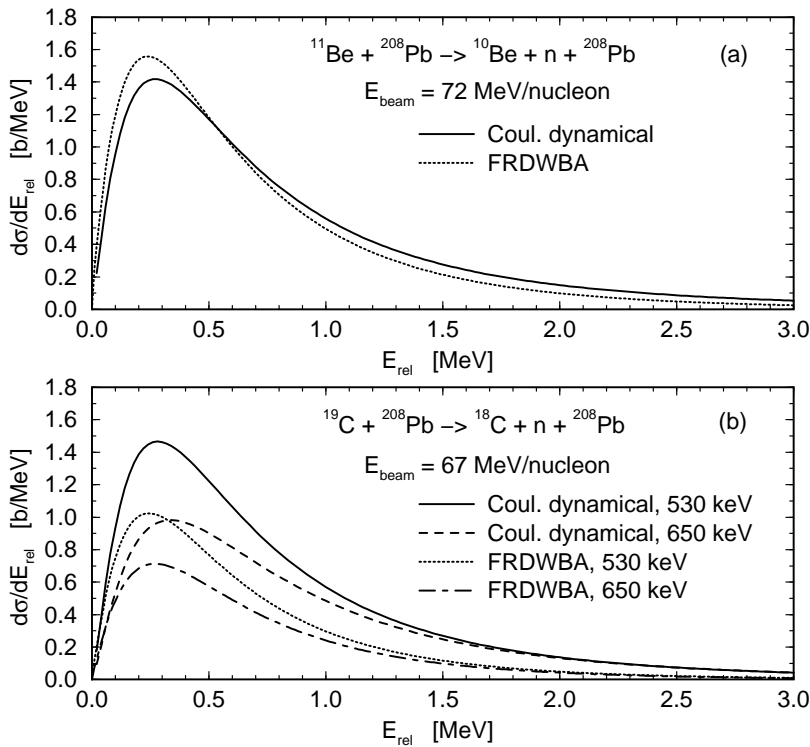,height=10.0cm}}
\end{center}
\caption{
(a) Comparison of the pure Coulomb dynamical model (solid line) and
the finite range DWBA (dotted line) contributions to the 
relative energy spectrum of the fragments (neutron and the core) 
emitted in the pure Coulomb breakup of $^{11}$Be on a $^{208}$Pb target
at the beam energy of 72 MeV/nucleon. 
(b) The same as in part (a) for the breakup of $^{19}$C on the 
same target at the beam energy of 67 MeV/nucleon. In this case the full
and dashed lines represent the results of the the dynamical model
and finite range DWBA, respectively, corresponding the $^{19}$C binding energy
of 530 keV while dashed and dashed-dotted lines show the same for  
the binding energy of 650 keV.
}
\label{fig:fige}
\end{figure}
\noindent
take into account effects which are beyond the semiclassical dynamical
model calculations. Therefore,
the results of the semiclassical first order perturbation theory will
differ from those of different higher order models in different ways.
Thus, the conclusion about the role of the higher
order effects will differ if the first order and higher order calculations
performed within two different theories are compared with each other.
It may also 
be remarked here that the spectroscopic factors extracted from the comparison
of the Coulomb breakup calculations of different models with the corresponding
data could be quite different from each other, as has already been pointed out
in \cite{ban00}.
 
\section{SUMMARY AND CONCLUSIONS}
In this paper, we have studied the Coulomb and the nuclear breakup of the
one-neutron halo nuclei, $^{11}$Be and $^{19}$C, on a $^{208}$Pb
target within a semiclassical fully dynamical model of the projectile
excitation process. The time evolution of the projectile system
is described by numerically solving the time dependent Schr\"odinger
equation in three dimensions. Through this nonperturbative method
higher order effects are fully taken into account in the breakup
process. Unlike previous such calculations, we assume the
projectile to move along a hyperbolic trajectory instead of a straight-line
one. A simple single particle potential model was used to calculate
the ground state wave function of the projectile nuclei. 

With a configuration for the $^{11}$Be ground state in which
a 2$s_{1/2}$ neutron is coupled to the ground state of the $^{10}$Be core
with a binding energy of 0.504 MeV and a spectroscopic factor of unity,
the experimental relative energy distributions of the fragments
(neutron and $^{10}$Be) emitted in the breakup reaction of $^{11}$Be
on a lead target at the beam energy of 72 MeV/nucleon are described
rather well by our model. The Coulomb breakup dominates these cross
sections around the peak region while the nuclear breakup is important
at higher relative energies. This provides a natural explanation for the
failure of the pure Coulomb breakup calculations in describing these data
at the larger relative energies. 

For the $^{19}$C case, the comparison of the dynamical calculations
with the data for the relative energy spectrum of the fragments suggests
that the ground state configuration of this nucleus is consistent with
a 2$s_{1/2}$ neutron coupled to the ground state of the $^{18}$C core with
a binding energy of 0.650 MeV and a spectroscopic factor of 1.
Alternatively, it could also have a binding energy of 0.530 MeV but
a spectroscopic factor of about $0.6-0.7$. The current data \cite{nak99} on
the relative energy spectrum are not sufficient to rule out either of
these possibilities.  Further experimental studies in
which the relative energy spectrum is measured by tagging the specific core
states and calculations of the Coulomb and nuclear breakup effects 
within different models (e.g., finite range DWBA) are clearly needed
to settle this issue.

By comparing the dynamical model results for the pure Coulomb breakup 
with the first order calculations (which are deduced from
the same model in a particular limit), we find that   
higher order effects in these reactions are generally small.
They reduce the magnitudes of the first order relative energy distribution
of the fragments by about 10-15$\%$ in the peak region. Their shapes,
however, remain almost unaltered by the higher order effects. However,
a comparison of the higher order and first order calculations of  
two different models may lead to large differences in the two results.
This is due to the fact that higher order calculations performed within
different theories include these effects in different 
approximations and could differ quite a bit from each other.

\section{ACKNOWLEDLEGMENTS}
One of us (R.S.) would like to thank Pawel Danielewicz for his kind 
hospitality in the theory group of the National Superconducting 
Cyclotron Laboratory at the Michigan State University. This work has
been supported by the National Science Foundation grants No. PHY-0070818,
PHY-0070911, and PHY-9528844.

\end{document}